\documentclass[
twocolumn,
superscriptaddress,
showpacs,
english,
floatfix]
{revtex4-2}

\usepackage{float}
\usepackage{subcaption}
\usepackage[T1]{fontenc}
\usepackage[utf8]{inputenc}
\usepackage{babel}
\usepackage{amsmath}
\usepackage{amssymb}
\usepackage{wasysym}
\usepackage{graphicx}
\usepackage{multirow}
\usepackage{gensymb}
\usepackage[dvipsnames]{xcolor}
\usepackage{microtype}
\usepackage{xcolor}
\usepackage{soul}
\usepackage[version=3]{mhchem} 
\usepackage{chemformula}[2014/04/08]
\usepackage{booktabs}
\usepackage{bm}
\usepackage{url}
\usepackage{microtype}
\usepackage{blindtext}
\usepackage{mathptmx}

\usepackage[linktocpage=true,
  colorlinks=true, 
  pdfborder={0 0 0},
  linkcolor=blue,
  citecolor=red,
  filecolor=yellow,
  urlcolor=blue,
  bookmarks,
  pdfauthor={},
]{hyperref}

\usepackage[capitalise]{cleveref}

\newcommand{\tc}{{T$_{\text{c}}$}}
\newcommand{\UniBasel}{Department\ of\ Physics,\ University\ of\ Basel,\ Klingelbergstrasse\ 82,\ CH-4056\ Basel,\ Switzerland}
\newcommand{\biospin}{Bruker\ BioSpin\ AG,\ Industriestrasse\ 26,\ 8117\ Fällanden,\ Switzerland}

\begin{document}

\title{Ternary Phase Diagram of Nitrogen Doped Lutetium Hydrides}
\author{Moritz Gubler}          \affiliation{\UniBasel}
\author{Marco Krummenacher}       \affiliation{\UniBasel}
\author{Jonas A.\ Finkler}       \affiliation{\UniBasel}
\author{Jos\'e A.\ Flores-Livas} \affiliation{\biospin}
\author{Stefan Goedecker}       \affiliation{\UniBasel}

\begin{abstract}
This paper presents the results of an extensive structural search of ternary solids  containing lutetium, nitrogen and hydrogen. 
Based on thousands of thermodynamically stable structures, available online, the convex hull of the formation enthalpies is constructed. 
To obtain the correct energetic ordering, the highly accurate RSCAN DFT functional is used in high quality all-electron calculations. 
In this way possible pseudopotential errors are eliminated.
A novel lutetium hydride structure (\ce{HLu2}) that is on the convex hull is found in our search.
An electron phonon analysis however shows that it is not a candidate structure for near ambient superconductivity.
Besides this structure, which appears to have been missed in previous searches, possibly due to different DFT methodologies, our results agree closely with the results of previously published structure search efforts. 
This shows, that the field of crystal structure prediction has matured to a state where independent methodologies produce consistent and reproducible results, underlining the trustworthiness of modern crystal structure predictions.
Hence it is quite unlikely that a structure, that would give rise 
within standard BCS theory to the superconducting properties, claimed to have been observed by \citet{room_temp_lu}, exists. 
This solidifies the evidence that no structure with conventional superconducting properties exists that could explain the experimental observation made by \citet{room_temp_lu}

\end{abstract}

\maketitle

\section{Introduction}\label{Sec:Int}
In their recent publication \citet{room_temp_lu} claim to have experimentally observed superconductivity in bulk nitrogen doped lutetium hydride (Lu-N-H) at a \tc{} of 294\,K and at a pressure of 1\,GPa. Since no detailed analysis of the structure, that is claimed to be superconductive at near ambient conditions, is given, an explanation of the mechanism that could lead to the observed superconductivity is missing. 
The mystery of the exact composition and structure of the putative superconductor has raised great interest into Lu-N-H structures throughout the entire materials science and solid state physics community.

The reaction of the community to the news of another room temperature superconductor from Dias and coworkers was prompt. 
Already a few days later, \citet{shan2023pressure} published their experimental study about pressure induced color changes in \ce{LuH2}. 
The observed color changes in the samples are similar to the ones presented in Ref.~\cite{room_temp_lu} but resistivity measurements showed no signs of superconductivity above 1.5\,K. 
One of the first theoretical studies on the Lu-N-H system was conducted by \citet{luh_convex_hull}. 
Their work also focused on lutetium hydrides. In order to investigate the convex hull of Lu-H the evolutionary structure prediction algorithm from the USPEX~\cite{uspex} package was used. 
Liu \textit{et al.}\ found \ce{LuH2} to be the most stable lutetium hydride and they conclude that the \ce{LuH2} is the parent structure when lutetium hydrides are doped with Nitrogen.
An overview of the Lu-N-H convex hull can be found in the recent work of \citet{lunh-search} were the they present the results of a detailed structure search at ambient pressure.
In the study of Ferreira \textit{et al.}, the configurational space of the ternary Lu-N-H structure was investigated thoroughly using the USPEX~\cite{uspex} evolutionary search method and the AIRSS~\cite{airss} random structure search method. 
In the evolutionary search with USPEX, Ferreira \textit{et al.}\ calculated energy and forces on the DFT level and in the random structure search with AIRSS, ephemeral data derived potentials~\cite{eddp} were used. An electron phonon analysis of the best candidate structures for room temperature superconductivity from Ferreira \textit{et al.}\ disagrees with the observation of near ambient superconductivity made by \citet{room_temp_lu}. 
Based on their results, Ferreira \textit{et al.}\ conclude that the observations made by \citet{room_temp_lu} cannot be explained with the electron phonon mechanism that describes conventional superconductivity.

Given that \citet{luh_convex_hull}, \citet{hilleke_23} and \citet{lunh-search} have investigated the configurational and compositional space of Lu-N-H thoroughly the excitement about the Lu-N-H superconductor has been damped considerably as the observations made by \citet{room_temp_lu} could not be explained using the current state of the art theoretical materials science methods. 

There are basically three options that explain this disagreement between theory and experiment:
\begin{itemize}
    \item \citet{room_temp_lu} observed unconventional superconductivity.
    \item There is an error in the experimental setup of \citet{room_temp_lu}.
    \item The correct structure was not found in all theoretical structure searches.
\end{itemize}

In this paper we present the results of an independent structure search in the ternary Lu-N-H phase diagram, ruling further out the last possibility that an important structure was overlooked. 
All presented final results were obtained with the the RSCAN functional~\cite{rscan}, which is widely considered to be the most accurate functional for cohesive energies. Well tested pseudo-potentials for this functional are however scarce. To eliminate any pseudo-potential errors we have therefore performed highly accurate all-electron calculations. 
Therefore, our results are expected to be more accurate than all other previous results.
The same approach has recently been used 
in a large scale structure search~\cite{gublermissing} for the putative carbonaceous sulphur hydrides superconductor~\cite{snider2020retracted}. 

Our results solidify the conclusions from the previous studies~\cite{luh_convex_hull,hilleke_23,lunh-search} that no conventionally superconducting structure can be found in the ternary Lu-N-H phase diagram. 

\begin{figure}
    \centering
    \includegraphics{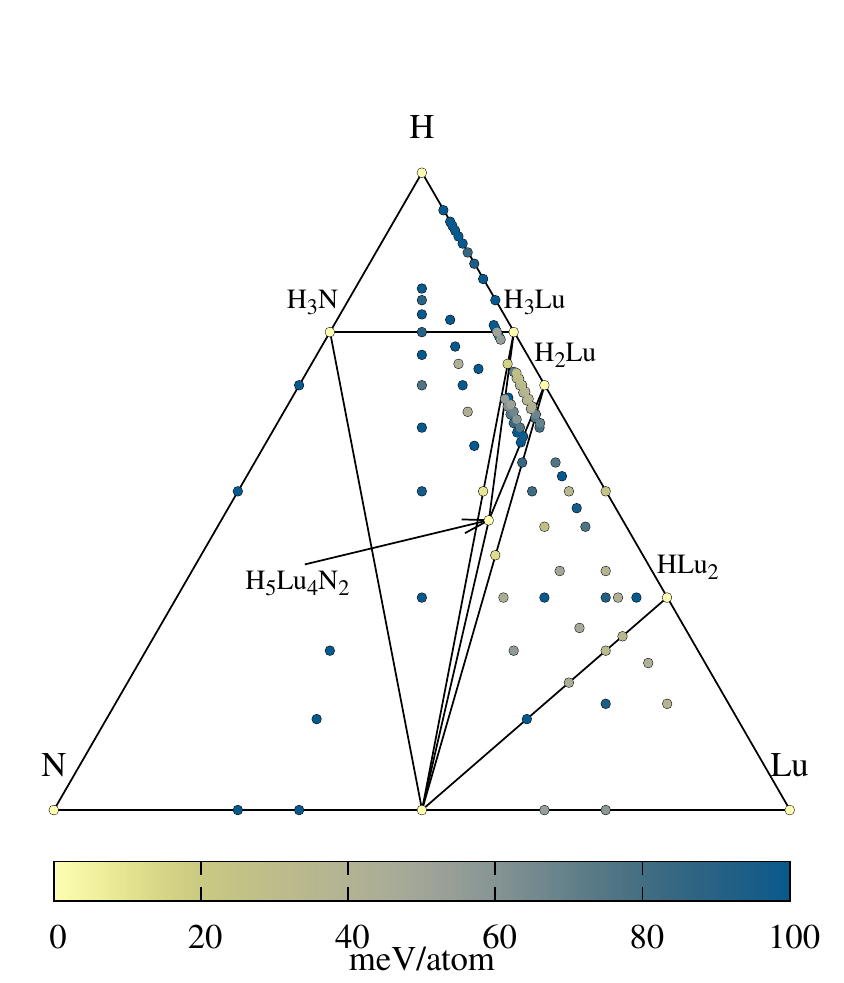}
    \caption{Formation enthalpy difference to the convex hull in meV per atom at 1\,GPa of pressure. The black lines indicate the convex hull.}
    \label{fig:convex_hull}
\end{figure}
\section{Methods}
\subsection{Structure search with Minima Hopping}\label{sec:minhop}
 The phase diagram of the ternary Lu-N-H structures were explored using the minima-hopping method~\cite{mh, Amsler2010, Sicher2011, Roy2008, sqnm}. 
Minima hopping is a method that reliably finds the global minimum of potential energy surfaces using a combination of variable cell shape molecular dynamics~\cite{parMD} along soft modes of the potential energy surface and variable cell shape geometry optimization~\cite{vcsqnm}. 
 Since it is not supposed to generate a thermodynamic distribution, it can escape from any funnel by crossing high energy barriers. 
 Because of that, minima hopping will always find the global minimum given a sufficiently long simulation. 
Other methods such as evolutionary search algorithms introduce moves to generate new structures which can be insufficient to escape from a deep funnel.

 In the Minima Hopping runs, energies, forces and the stress tensor were calculated on the DFT level with the standard PBE functional~\cite{sirius_pbe} and the SIRIUS library~\cite{sirius,sirius_lapw} which is a GPU accelerated and MPI parallelized plane wave code. 
 Ultrasoft pseudo potentials~\cite{DALCORSO2014} were used to eliminate the core electrons. 
 A plane wave cutoff of 1400\,eV was used and a tight $4 \times 4 \times 4$ Monkhorst-Pack~\cite{Monkhorst} k-point grid was chosen.

 In total 108 different stoichiometries were sampled at a pressure of 1\,GPa. 
 To ensure convergence of the minima hopping method, the search was only stopped after 25000 distinct local minima were found. 
 On average 230 different minima were found for every stoichiometry.
\begin{figure}
    \centering
    \includegraphics{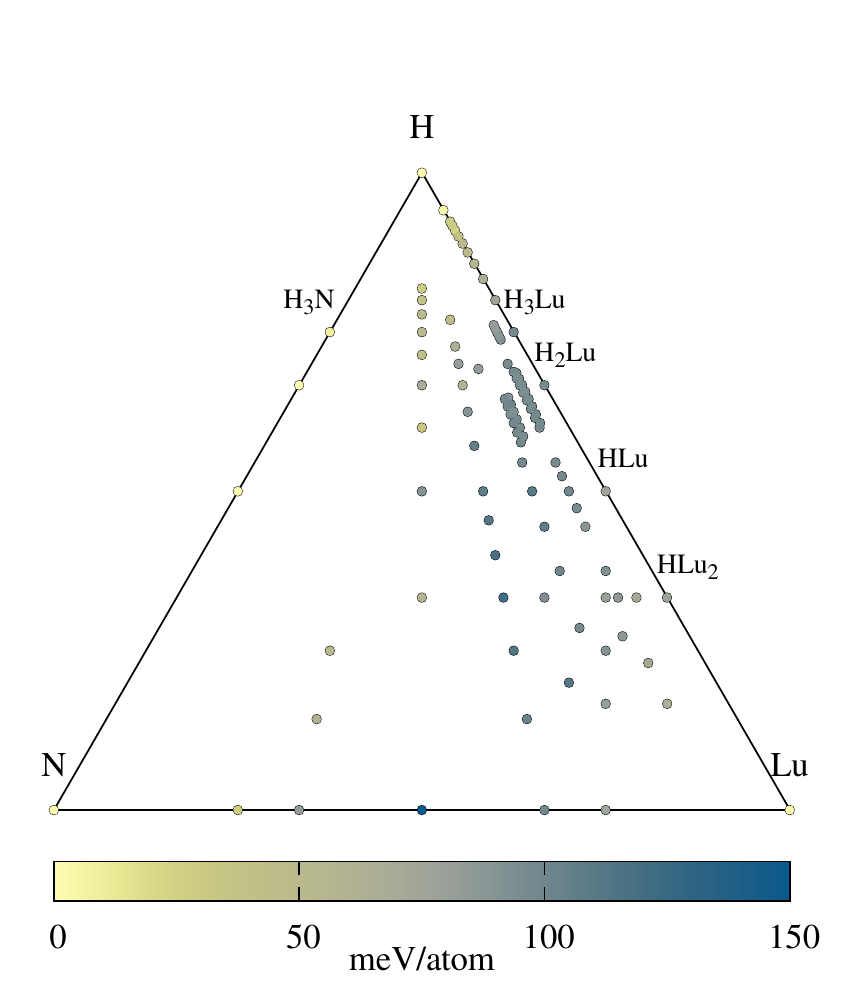}
    \caption{Difference in formation enthalpy for the ground state of each stoichiometry between a PBE pseudopotential calculation and an all electron RSCAN calculation at 1\,GPa of pressure.}
    \label{fig:compare}
\end{figure}
\subsection{All electron calculations}
In order to increase the accuracy of the DFT calculations from \cref{sec:minhop} the 20 lowest minima that are in the energy range of 50\,meV per atom compared to the ground state of each stoichiometry were recalculated with a more precise DFT method. 
With these criteria, 1600 structures were selected for further processing. The error introduced by the pseudopotentials was eliminated by performing an all electron calculation and the error from the PBE functional was reduced by using the accurate RSCAN exchange correlation functional~\cite{rscan, scan_accuracy_1,scan_accuracy_2}.

FHI-aims~\cite{fhi,fhi_grid,fhi_stress} was used for a geometry optimization of the 1200 systems with the previously mentioned settings, a $5 \times 5 \times 5$ $\Gamma$ centered k-point grid and the tier 2 basis set. The resulting energies are our most accurate ones.  They reduce errors in the the energetic ordering of the 1200 lowest systems and therefore the chance of finding the wrong ground state. All energies used in the convex hull plots of this paper were obtained using this procedure.

\section{Results}

\subsection{Comparison between the high-performance and high-accuracy DFT calculations}\label{sec:high_acc}
The energies of the most promising structures that were found using the minima hopping method were recalculated with all electron DFT simulations that used the RSCAN functional. The difference in formation enthalpy of the plane wave calculations with PBE functional and the all electron calculation with the RSCAN functional is displayed in \cref{fig:compare}. For most stoichiometries, the error is between 50\,meV per atom and 100\,meV per atoms. The energy difference between the pseudopotential and the all electron calculation is rather large. Nevertheless, errors of this magnitude are not too uncommon in DFT calculations.
The good correlation in formation enthalpy of the nitrogen hydrides indicates that lutetium is responsible for a large part of the energy error.

Even though the energetic error of the pseudopotential calculations is rather large, the energetic ordering of the structures on the convex hull is surprisingly good and the ground state structures were predicted correctly by the pseudopotential calculations.

\subsection{Lutetium hydrides}
\begin{figure}
    \centering
    \includegraphics{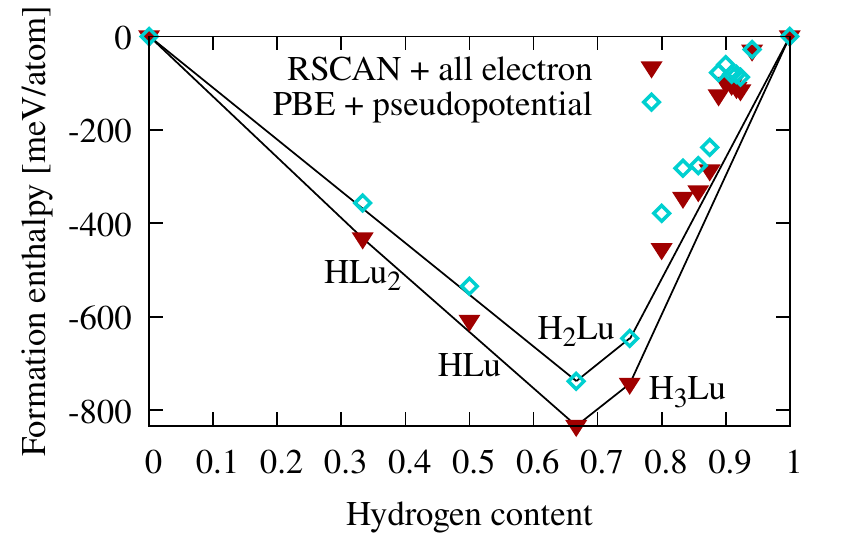}
    \caption{A comparison of the binary convex hull of the Lu-H system calculated with the high throughput and high accuracy DFT methods. The convex hull is displayed by the black lines.}
    \label{fig:binary}
\end{figure}

In order to get an initial impression for the Lu-N-H system, a structure search for binary lutetium hydrides was first conducted and the formation enthalpies were also verified using highly accurate all electron calculations. The convex hull of the binary system is displayed in \cref{fig:binary}. \ce{H3Lu}, \ce{H2Lu} and \ce{HLu2} all lie on the convex hull of formation enthalpies which makes them thermodynamically stable. \ce{HLu} is only 16\,meV per atom above the convex hull which is within the typical uncertainty of DFT calculations. Therefore, \ce{HLu} may also be thermodynamically stable.
The high accuracy DFT calculations presented in \cref{sec:high_acc} show that \ce{HLu2} is on the convex hull. The structure and the electronic density of states of \ce{H3Lu}, \ce{H2Lu} and \ce{HLu2} is displayed in \cref{fig:h2lu,fig:h3lu,fig:hlu2,fig:dos}. Because neither \citet{luh_convex_hull} or \citet{lunh-search} have found an \ce{HLu2} structure that is on the convex hull we decided to calculate the \tc{} using Quantum Espresso\cite{qe-2009} and the Allen-Dynes equation~\cite{allen-dynes-75}. The calculations were done using the same procedure as in Ref.~\cite{gublermissing}. To calculate the \tc{} of \ce{HLu2}, a plane wave cutoff 1370\,eV, a $16 \times 16 \times 16$ k-grid (a spacing of $\sim 0.03 \,\text{A}^{-1}$) and a $4 \times 4 \times 4$ q-grid (a spacing of $\sim 0.13\,\text{A}^{-1}$) were used. The resulting \tc{} is 0.5\,K which is obtained using the value 0.1 for $\mu^*$.

\subsection{Ternary lutetium nitrogen hydrides}
\begin{figure*}
\begin{subfigure}{.5\textwidth}
      \includegraphics[width=.9\linewidth]{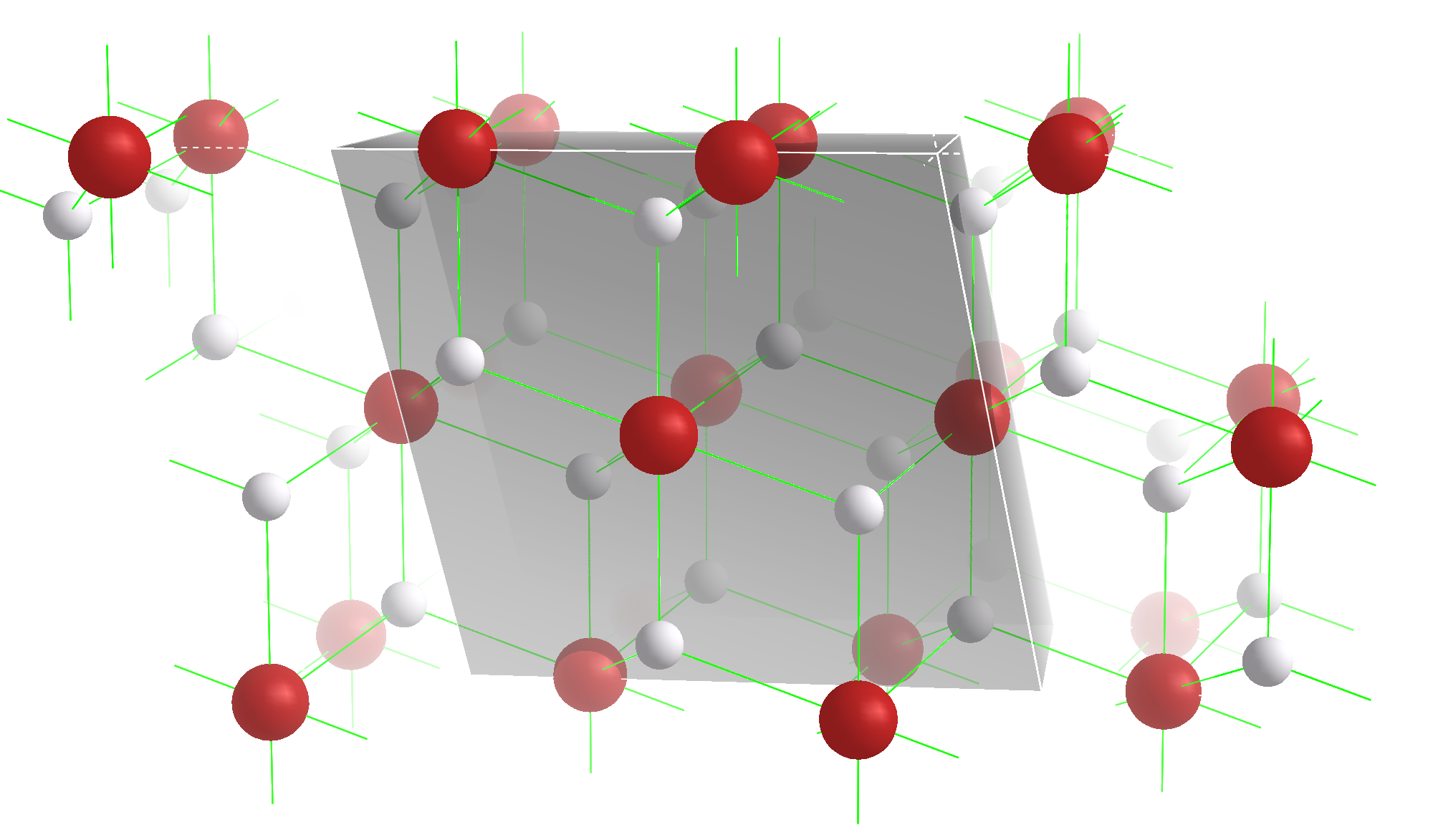}
    \caption{\ce{H2Lu}}
    \label{fig:h2lu}
\end{subfigure}%
\begin{subfigure}{.5\textwidth}
      \includegraphics[width=.9\linewidth]{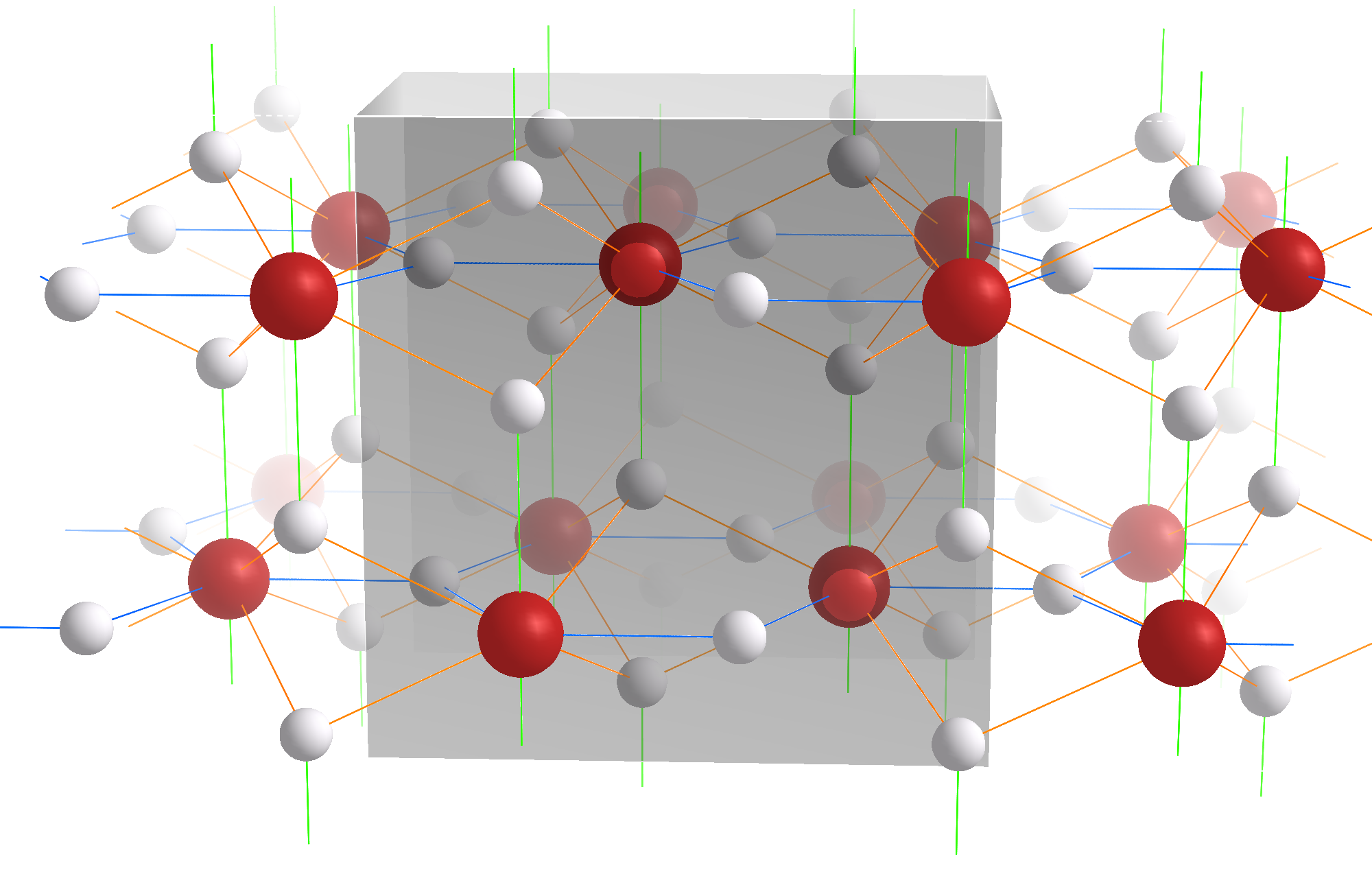}
    \caption{\ce{H3Lu}}
    \label{fig:h3lu}
\end{subfigure}
\begin{subfigure}{.5\textwidth}
      \includegraphics[width=.9\linewidth]{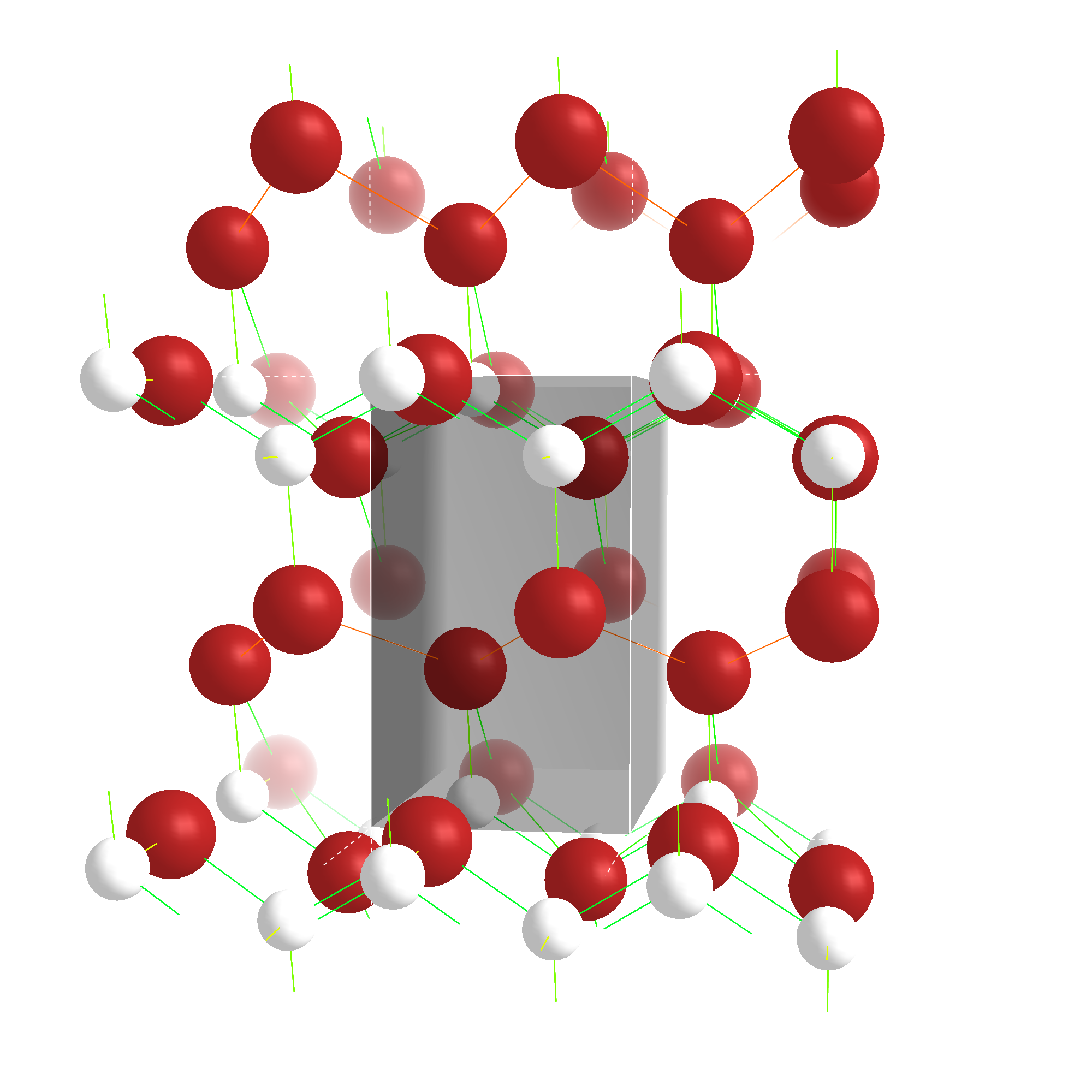}
    \caption{\ce{HLu2}}
    \label{fig:hlu2}
\end{subfigure}%
\begin{subfigure}{.5\textwidth}
      \includegraphics[width=.9\linewidth]{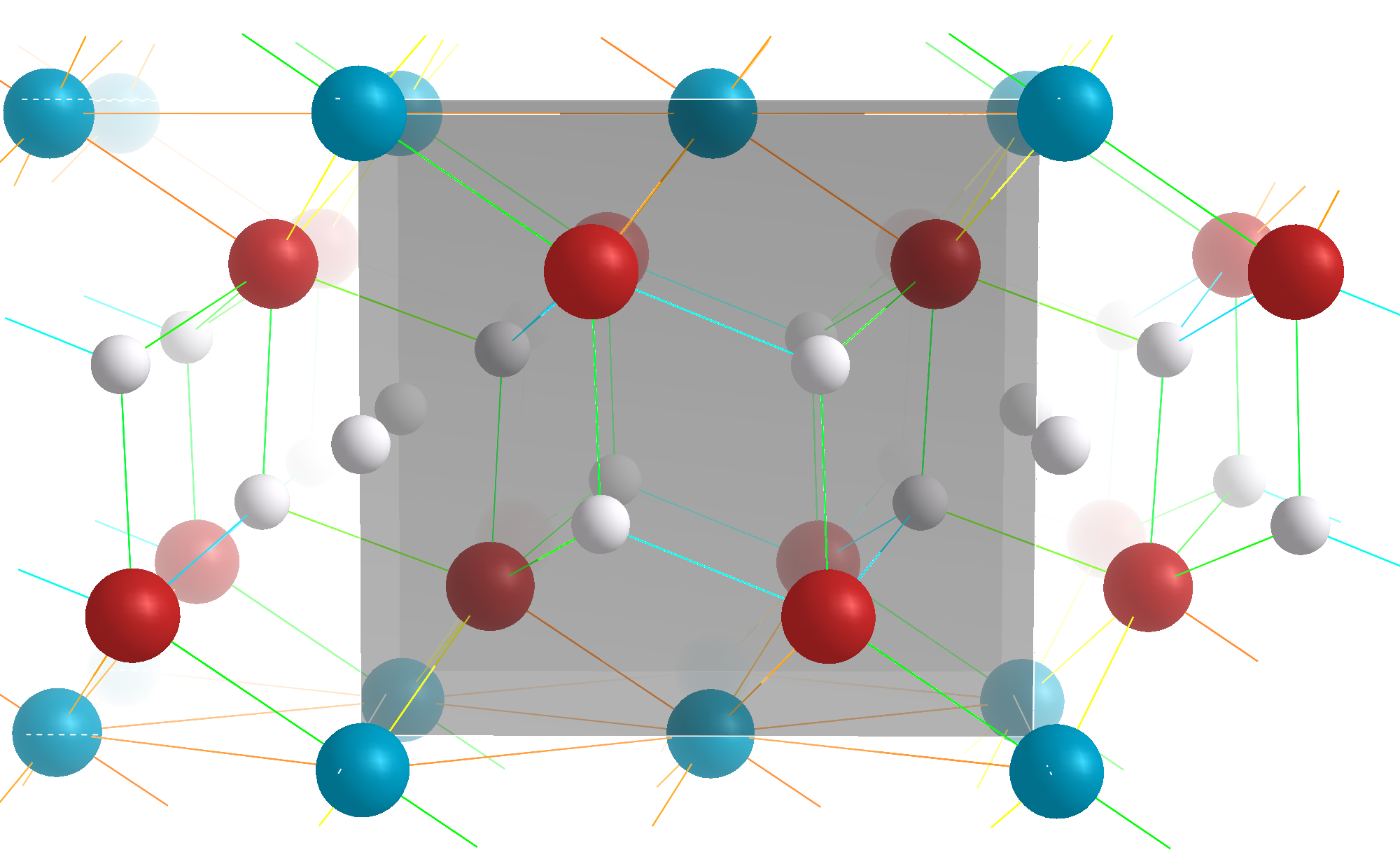}
    \caption{\ce{H5Lu4N2}}
    \label{fig:h5lu4n2}
\end{subfigure}
\begin{subfigure}{\textwidth}
    \includegraphics{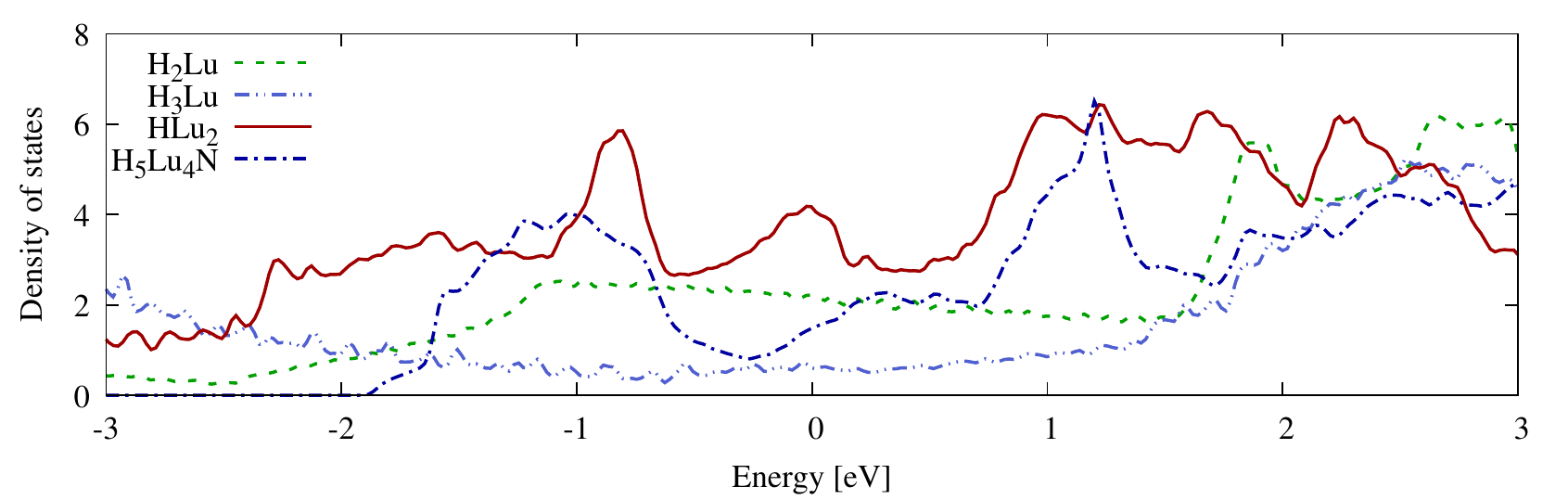}
    \caption{Electronic density of states of the structures from \cref{fig:h2lu,fig:h3lu,fig:hlu2,fig:h5lu4n2}}
    \label{fig:dos}
\end{subfigure}
\caption{A selection of structures that lie on the convex hull at 1\,GPa of pressure and their electronic density of state calculated with an all electron RSCAN simulation. Hydrogen is pictured in white, nitrogen in blue and lutetium in red and the Fermi level is shifted to zero.}
\label{fig:structures}
\end{figure*}

There is only one ternary stoichiometry, that is on the convex hull: \ce{H5Lu4N2}. 
It is pictured in \cref{fig:h5lu4n2} and a plot of the electronic density of states can be found in \cref{fig:dos}. 
Our results indicate that nitrogen doped \ce{LuH_x} crystals are thermodynamically unstable. 
A detailed view of the convex hull can be seen in \cref{fig:convex_hull}.

\section{Conclusion}
Theoretical structure prediction is by no means an easy or routine task, especially for complex ternary systems such as Lu-N-H involving rare elements such as \ce{Lu}. 
It is therefore reassuring, that two independent studies based on completely different methodologies and codes come to comparable conclusion.
The only notable difference between our study and the previous studies is that we have identified \ce{HLu2} to be on the convex hull. Our calculation of the \tc of \ce{HLu2} shows that it is not responsible for the high \tc measured by \citet{room_temp_lu}
Otherwise, very similar structures were found which lead to a comparable convex hull. 
Only a few (\ce{H2Lu}, \ce{H3Lu}, \ce{HLu2}, \ce{LuN} and \ce{H5Lu4N2}) Lu-N-H stoichiometries lie on this convex hull at 1\,GPa of pressure. This is consistent with prior structure searches~\cite{luh_convex_hull,hilleke_23,lunh-search}. 
Firstly, this highlights the fact, that modern crystal structure prediction methods have reached a level of maturity, where consistent and reproducible results can be expected. 
Secondly, the fact that two studies, based on an independent methodologies, come to the same central conclusion reduces the risk that a bias was introduced during the structure search and that a possible structure with superconducting properties was overlooked. 
This allows us to conclude, with a very high certainty, that no conventionally superconducting Lu-N-H structure exists.

\section{Structural data}
Low enthalpy Lu-N-H structures can be found in this GitHub repository: \url{https://github.com/moritzgubler/H-Lu-N}. The enthalpy of these structures was calculated as described in \cref{sec:high_acc}.

\begin{acknowledgments} 
The calculations were performed on the computational resources of the Swiss National Supercomputer (CSCS) under project s1167 and on the Scicore (\url{http://scicore.unibas.ch/}) computing center of the University of Basel. Financial support was obtained from the Swiss National Science Foundation, project $200021\_191994$.
\end{acknowledgments}
\newpage
\bibliographystyle{apsrev4-1}
\bibliography{main}

\end{document}